\documentclass{aastex631}
\usepackage[table]{xcolor}

\received{October 15, 2025}
\revised{November 3, 2025}
\accepted{January 16, 2026}

\begin{document}

\title{Unveiling Hidden Clustering: An Unsupervised Machine Learning
Study of Repeating FRB 20220912A}

\author[0009-0006-0970-8568]{An-Chieh Hsu}
\affiliation{Department of Physics, National Chung Hsing University, 145 Xingda Rd., South Dist., Taichung 40227, Taiwan}
\correspondingauthor{An-Chieh Hsu}
\email{angel20040212@gmail.com}

\author[0000-0001-7228-1428]{Tetsuya Hashimoto}
\affiliation{Department of Physics, National Chung Hsing University, 145 Xingda Rd., South Dist., Taichung 40227, Taiwan}

\author[0000-0002-6821-8669]{Tomotsugu Goto}
\affiliation{Department of Physics, National Tsing Hua University, 101, Section 2. Kuang-Fu Road, Hsinchu, 30013, Taiwan}

\affiliation{Institute of Astronomy, National Tsing Hua University, 101, Section 2. Kuang-Fu Road, Hsinchu, 30013, Taiwan}

\author[0000-0001-6010-714X]{Tomoki Wada}
\affiliation{Frontier Research Institute for Interdisciplinary Sciences, Tohoku University, Sendai, Japan}
\affiliation{Astronomical Institute, Graduate School of Science, Tohoku University, Sendai, Japan}
\affiliation{Department of Physics, National Chung Hsing University, 145 Xingda Rd., South Dist., Taichung 40227, Taiwan}

\author[0000-0003-0054-6081]{Bjorn Jasper Raquel}
\affiliation{Theoretical Physics Group, National Institute of Physics, University of the Philippines, Diliman, Quezon City 1101, Philippines}




\begin{abstract}
Fast Radio Bursts (FRBs) are millisecond-duration radio transients of extragalactic origin. Classifying repeating FRBs is essential for understanding their emission mechanisms, but remains challenging due to their short durations, high variability, and increasing data volume. Traditional methods often rely on subjective criteria and struggle with high-dimensional data.
In this study, we apply an unsupervised machine learning framework—combining Uniform Manifold Approximation and Projection (UMAP) and Hierarchical Density-Based Spatial Clustering of Applications with Noise (HDBSCAN)—to eight observed parameters from FRB 20220912A. Our analysis reveals three distinct clusters of bursts with varying spectral and fluence properties.
Comparisons with clustering studies on other repeaters, show that some of our clusters share similar features, such as FRB 20201124A and FRB121102, suggesting possible common emission mechanisms. We also provide qualitative interpretations for each cluster, highlighting the spectral diversity within a single source. Notably, one cluster shows broadband emission and high fluence, typically seen in non-repeating FRBs, raising the possibility that some non-repeaters may be misclassified repeaters due to observational limitations. Our results demonstrate the utility of machine learning in uncovering intrinsic diversity in FRB emission and provide a foundation for future classification studies.

\end{abstract}

\keywords{Radio transient sources ---  Time domain astronomy --- Radio bursts}


\section{Introduction} \label{sec:intro}

Fast Radio Bursts (FRBs) are millisecond-duration radio transients of extragalactic origin, first discovered by  \cite{2007Sci...318..777L}. 
Their high dispersion measures (DMs) indicate propagation through substantial intergalactic distances \citep[e.g.,][]{2017arXiv170902189P}. FRBs are broadly classified into two categories: repeating FRBs, which emit multiple bursts over time, and non-repeating FRBs, which have only been observed to emit a single burst \citep[e.g.,][]{2019ARA&A..57..417C}. Distinguishing between these two classes is critical for understanding the underlying astrophysical mechanisms responsible for their emission.


Among the recently identified repeaters, FRB 20220912A has drawn considerable interest since its discovery by CHIME/FRB on 12 September 2022 \citep{2022ATel15679....1M}. FAST observations during its active phase detected a large number of bursts, covering frequencies from 800 to 1600 MHz \citep{2023ApJ...955..142Z}. These bursts exhibit narrow-band spectra, diverse morphologies, and clear scintillation signatures, providing valuable probes of both the emission process and the intervening plasma \citep[e.g.,][]{2024SCPMA..6719512W}. The high burst rate observed with FAST makes FRB 20220912A one of the most active known repeaters, offering an important laboratory for testing theoretical models of repeating FRBs.

Despite these advancements, classifying bursts from repeating FRBs remains challenging. For FRB 20220912A, bursts exhibit empirical correlations in properties such as fluence and width, and their energy distribution has been found to follow a log-normal form \citep{2023ApJ...955..142Z}. Additionally, multi-wavelength observations reveal that the cumulative spectral energy distribution follows a power-law trend \citep{2024A&A...690A.219P}. These statistical characterizations provide a basis for understanding the diversity of burst properties, though the underlying physical mechanisms remain under investigation.               

Understanding these properties is crucial for developing classification schemes, yet previous attempts at FRB classification have often relied on a limited set of observational parameters with some physical basis. However, current understanding of the FRB radiation mechanism is far from ideal. As a result, such approaches are susceptible to human biases and may overlook important patterns within the data.

Machine learning offers a data-driven alternative, capable of handling multiple physical parameters simultaneously and mitigating subjective biases. In particular, unsupervised learning methods such as Uniform Manifold Approximation and Projection (UMAP) and Hierarchical Density-Based Spatial Clustering of Applications with Noise (HDBSCAN) have recently been applied to FRB data to identify potential subtypes based on intrinsic properties \citep{2023MNRAS.521.5738C, 2025arXiv250618854J, 2025ApJ...980..185S}. These techniques enable the discovery of latent structures within the FRB population and may provide new insights into their physical origins.

In this study, we aim to extend these efforts by applying an unsupervised machine learning framework to classify bursts from known FRB 20220912A. 
We focus on identifying physically meaningful subclusters that may correspond to different emission mechanisms, the geometry of the emission region, and/or propagation effects. 

This paper is organized as follows: we introduce the data selection used for the UMAP projection in Section~\ref{sec:data-preprocessing}. The methodology of our machine learning model is presented in Section~\ref{sec:methodology}. The results and analysis of the embedding clusters are shown in Section~\ref{sec:result}. More physical explanations of each cluster and comparisons with previous work are provided in Section~\ref{sec:discussion}, followed by the overall conclusion in Section~\ref{sec:conclusion}.

\section{Data Preprocessing}\label{sec:data-preprocessing}

We use the FRB catalogue detected using the Five-hundred-meter Aperture Spherical radio Telescope (FAST).
The catalogue includes a total of 1076 FRBs from the identical source of FRB 20220912A \citep{2023ApJ...955..142Z}.
The catalogue provides various observational parameters, including burst arrival time (MJD), dispersion measure (pc $\cdot$ cm$^{-3}$), peak flux (mJy), width (ms), peak frequency (MHz), bandwidth (MHz), fluence (Jy $\cdot$ ms), energy (10$^{36}$~erg), rotation measure (rad $\cdot$ m$^{-2}$), and degrees of linear and circular polarization (\%).

We aim to preserve as many observational parameters as possible. However, we exclude the Burst Arrival Time(MJD), as it does not reflect the intrinsic properties of FRBs, and the Dispersion Measure (pc $\cdot$ cm$^{-3}$), as it is more related to the environment rather than the origin of the FRB. 
Using the Burst Arrival Time (MJD) provided by the catalog, we introduced a new parameter, Waiting Time, defined as the time difference between consecutive bursts, specifically the arrival time of the following burst minus that of the current burst. 
Bursts with a Waiting Time greater than 0.5 days were excluded because they are affected by the daily monitoring period.
Following \citet{2023ApJ...955..142Z}, we excluded bursts with a peak frequency outside the range of 1–1.5 GHz to mitigate a possible bias introduced by the limited bandwidth of the telescope. 
Bursts with a peak frequency within 50 MHz of the bandwidth edges (1050–1450 MHz) and a bandwidth less than 100 MHz were also excluded because of the large uncertainty in the spectral fitting of the dynamic spectra \citep{2023ApJ...955..142Z}. 
We also excluded Peak Flux (mJy) and Energy (10$^{36}$erg), as they are not directly observable but are instead derived from Fluence (Jy $\cdot$ ms) and describe almost the same physical quantity.

To remove a potential bias caused by different dynamic ranges of the input features, we applied data standardization using the \texttt{StandardScaler} from \texttt{sklearn.preprocessing}. 
This transformation adjusts all features to have a mean of 0 and a standard deviation of 1, preventing features with larger scales from dominating the UMAP analysis and improving the fairness of the clustering process.

After applying these criteria and transformations, the final dataset for UMAP analysis consists of 817 bursts. 
The parameters used for UMAP include:

\begin{itemize}
  \item Width (ms)
  \item PeakFrequency (MHz)
  \item BandWidth (MHz)
  \item Fluence (Jy $\cdot$ ms)
  \item Rotation Measure (rad $\cdot$ m$^{-2}$)
  \item Linear Polarization (\%)
  \item Circular Polarization (\%)
  \item Waiting Time (s)
\end{itemize}

\section{Data Processing/ Methodology} \label{sec:methodology}

\subsection{Uniform Manifold Approximation and Projection (UMAP)} 
UMAP is a widely used dimensionality reduction technique that facilitates the visualization and analysis of high-dimensional data. 
It works by constructing a low-dimensional representation of the data while preserving its local and global structures. 
After preprocessing, our dataset contains 817 rows and 8 columns, which serve as input for UMAP training.

UMAP has several important hyperparameters, among which the most critical is \texttt{n\_neighbors}. 
This parameter determines how UMAP balances the preservation of local versus global data structures. 
Specifically, it defines the size of the local neighborhood considered when learning the manifold structure of the data. 
Lower values of \texttt{n\_neighbors} prioritize capturing local structure, while higher values emphasize global relationships by incorporating more distant neighborhoods. 

To optimize the critical hyperparameter \texttt{n\_neighbors}, we evaluated the clustering performance using the \texttt{Silhouette Score} \citep{ROUSSEEUW198753} and \texttt{Davies-Bouldin Score} \citep{4766909}. 
The \texttt{Silhouette Score} quantifies how well each data point fits within its assigned cluster compared to the closest neighboring cluster, with higher values indicating better-defined clusters. 
Conversely, the \texttt{Davies-Bouldin Score} evaluates the average similarity ratio between each cluster and its most similar cluster, where lower values signify better clustering performance.

\begin{figure}[htbp]
    \centering
    \includegraphics[width=\textwidth]{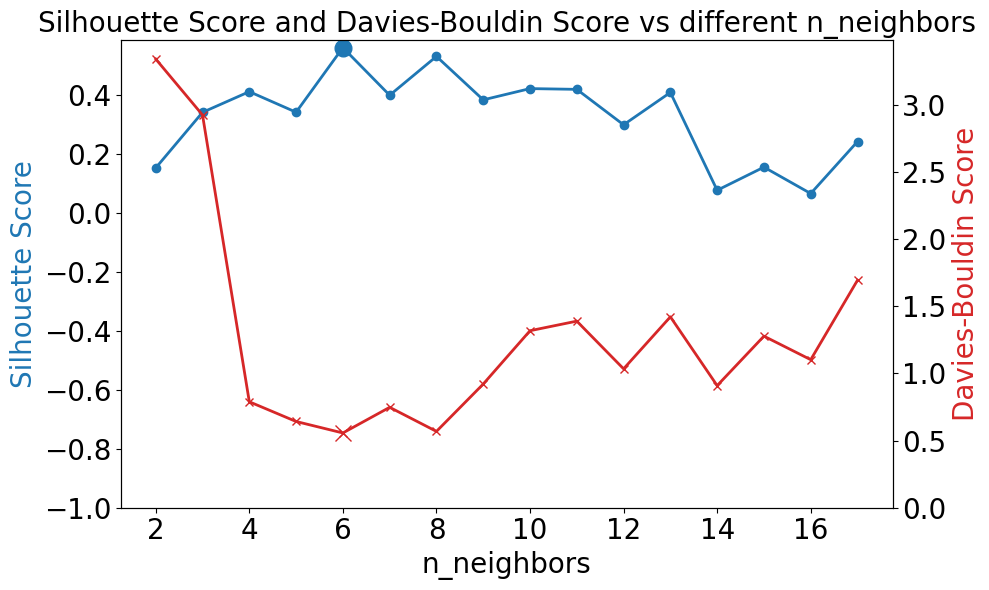}
    \caption{The \texttt{Silhouette Score} and \texttt{Davies-Bouldin Score} for different values of \texttt{n\_neighbors}. For clustering, a higher \texttt{Silhouette Score} and a lower \texttt{Davies-Bouldin Score} indicate better performance. As shown in this figure, when \texttt{n\_neighbors} = 6, the \texttt{Silhouette Score} reaches its maximum and the \texttt{Davies-Bouldin Score} reaches its minimum. Therefore, we choose \texttt{n\_neighbors} = 6 for the following analysis.}
    \label{fig:myfigure-Silhoutte-Score}
\end{figure}

As shown in Figure~\ref{fig:myfigure-Silhoutte-Score}, when the hyperparameter \texttt{n\_neighbors} is set to 6, the \texttt{Silhouette Score} is maximized and the \texttt{Davies–Bouldin Score} is minimized, indicating optimal clustering quality. Another key decision is the dimensionality of the reduced representation. For visualization purposes, we chose \texttt{n\_components} = 2, which projects the data onto a two-dimensional plane, making it easier to interpret.

Finally, the parameter \texttt{min\_dist} controls the minimum spacing between points in the reduced space. 
Smaller values of \texttt{min\_dist} result in denser clusters, while larger values spread the points more evenly. 
For our research objectives, we directly adopted \texttt{min\_dist} = 0 from \cite{2023MNRAS.524.1668R} to balance cluster compactness and interpretability.

\subsection{Hierarchical Density-Based Spatial Clustering of Applications with Noise (HDBSCAN)} 

HDBSCAN is an unsupervised clustering algorithm designed to identify clusters of varying density while effectively handling noise and outliers in the data.
Unlike traditional clustering methods, HDBSCAN does not require a predefined number of clusters; instead, it discovers clusters based on the density distribution of data points. 
After projecting the data onto a 2D plane using UMAP, we applied HDBSCAN to cluster the resulting projection.

Similar to UMAP, HDBSCAN has several important hyperparameters. Among them, the most critical is the \texttt{min\_cluster\_size} parameter, which specifies the minimum number of data points required to form a cluster. 
For our analysis, we set \texttt{min\_cluster\_size} = 100, ensuring that only sufficiently large groups of points are considered as clusters.

Another key parameter is the \texttt{min\_samples} setting, which defines the minimum number of neighboring points required for a point to be classified as a core point. 
A larger value leads to more stringent clustering, with more points being treated as noise or outliers. 
In our study, we set \texttt{min\_samples} to 10 to balance the identification of clusters and the exclusion of noise, following the approach of \cite{2023MNRAS.524.1668R}.

\subsection{Sensitivity Analysis}
To evaluate the robustness of the selected UMAP–HDBSCAN hyperparameters, we performed a local sensitivity analysis. As described earlier, our baseline configuration uses UMAP with
\texttt{n\_neighbors} = 6 and \texttt{min\_dist} = 0, and HDBSCAN with \texttt{min\_cluster\_size} = 100 and \texttt{min\_samples} = 10. In the sensitivity test, UMAP’s \texttt{n\_neighbors} was varied from 4 to 8 and \texttt{min\_dist} from 0.0 to 0.1, while HDBSCAN’s \texttt{min\_cluster\_size} was varied from 80 to 120 and \texttt{min\_samples} from 8 to 12. For each hyperparameter combination, we repeated the clustering over 30 random seeds and quantified the stability of (i) the number of clusters, (ii) the noise fraction, and (iii) the Adjusted Rand Index (ARI) relative to the baseline result.

\subsection{Feature Selection and Waiting Time Control}
To assess the potential influence of observational cadence, we repeated the clustering analysis without including the Waiting Time feature, even though bursts with Waiting Time greater than 0.5 days had already been excluded in the original analysis. The optimal UMAP hyperparameter \texttt{n\_neighbors} was re-selected based on the \texttt{Silhouette Score} and \texttt{Davies-Bouldin} Score. This procedure allows us to verify whether the three-cluster structure persists independently of the Waiting Time feature.

\subsection{Alternative Dimensionality Reduction and Clustering Approaches}
\begin{figure}[htbp]
    \centering
    \includegraphics[width=\textwidth]{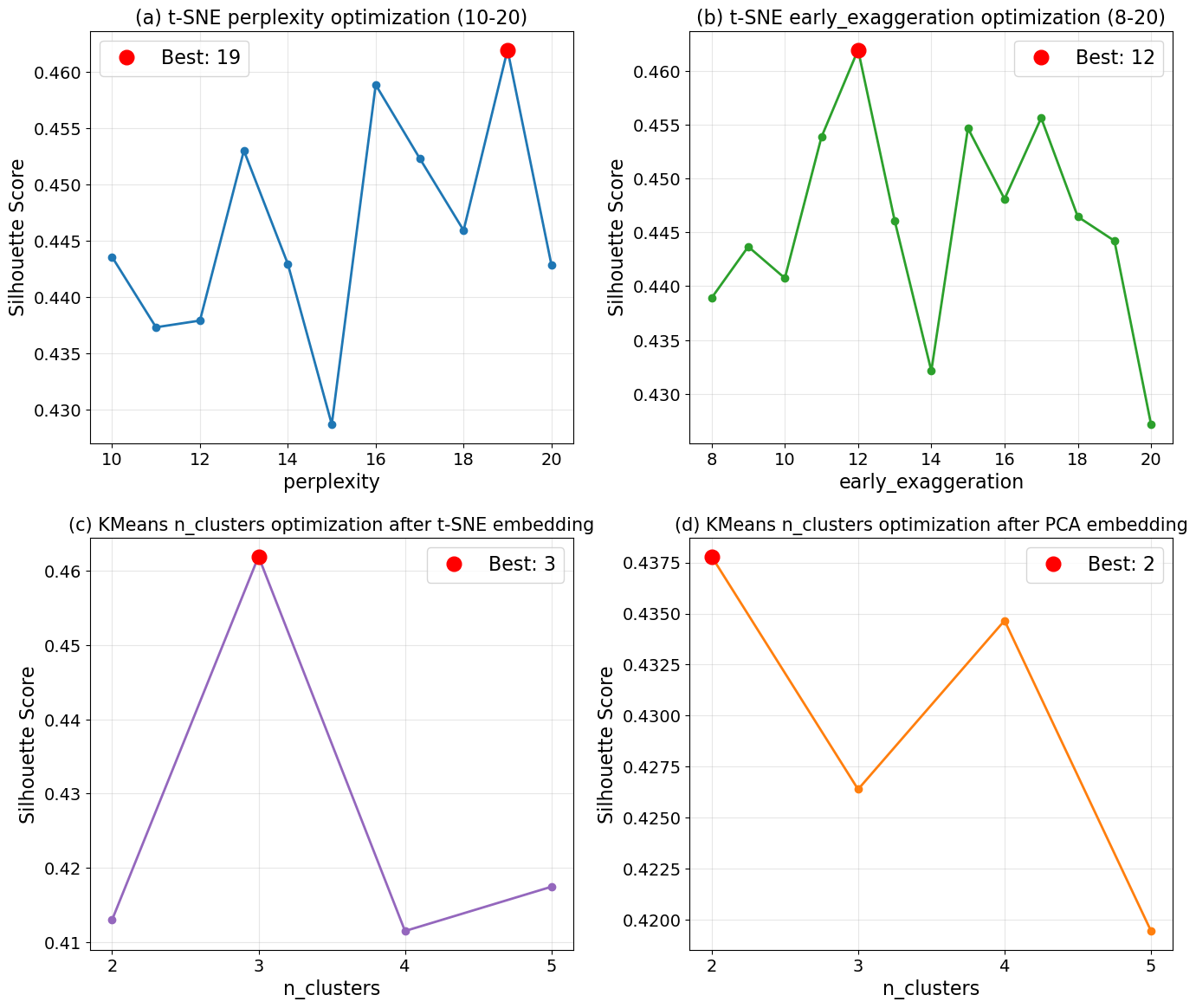}
    \caption{(a) \texttt{Silhouette Scores} as a function of t-SNE \texttt{perplexity}, maximized at \texttt{perplexity = 19}. (b) \texttt{Silhouette Scores} as a function of t-SNE \texttt{early\_exaggeration}, maximized at \texttt{early\_exaggeration = 12}. (c) \texttt{Silhouette Scores} as a function of KMeans \texttt{n\_clusters} after t-SNE embedding, maximized at \texttt{n\_clusters = 3}. (d) \texttt{Silhouette Scores} as a function of KMeans \texttt{n\_clusters} after PCA embedding, maximized at \texttt{n\_clusters = 2}.}
    \label{fig:myfigure-hyperparameter-optimization}
\end{figure}

To assess whether the identification of three clusters depends on the choice of dimensionality reduction and clustering methods, we performed additional analyses using three alternative approaches: Principal components analysis (PCA) \citep{2014arXiv1404.1100S} followed by K-means Clustering (KMeans) \citep{8ddb7f85-9a8c-3829-b04e-0476a67eb0fd}, PCA followed by HDBSCAN, t-Distributed Stochastic Neighbor Embedding (t-SNE) \citep{JMLR:v9:vandermaaten08a} followed by KMeans.  For both PCA- and t-SNE-based KMeans, the number of clusters (\texttt{n\_clusters}) was optimized using the \texttt{Silhouette Score}. For t-SNE itself, the hyperparameters \texttt{perplexity} and \texttt{early\_exaggeration} were also optimized using the \texttt{Silhouette Score} (see Fig.~\ref{fig:myfigure-hyperparameter-optimization}).

\subsection{Software Environment}
All analyses were performed in Python 3.12.3. The following Python libraries were used: numpy 1.26.4 \citep{2021ATel14516....1K} for numerical computations; pandas 2.2.2 \citep{reback2020pandas, mckinney-proc-scipy-2010} for data handling and table management; scikit-learn 1.5.1 \citep{scikit-learn} for preprocessing, dimensionality reduction (PCA, t-SNE), clustering (KMeans), and evaluation metrics (\texttt{Silhouette Score} and \texttt{Davies–Bouldin Score}); umap-learn 0.5.6 \citep{mcinnes2018umap-software} for nonlinear dimensionality reduction; HDBSCAN 0.8.37 \citep{McInnes2017} for density-based clustering; and matplotlib 3.8.4 \citep{Hunter:2007} and seaborn 0.13.2 \citep{Waskom2021} for data visualization.

\section{Result} \label{sec:result}
\begin{figure}[htbp]
    \centering
    \includegraphics[width=\textwidth]{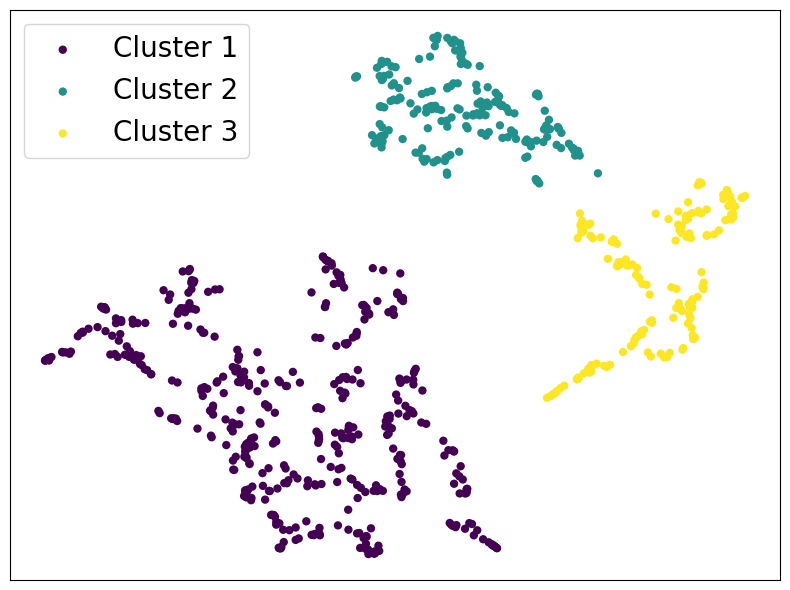}
    \caption{The unsupervised machine learning analysis of FRB 20220912A also yields three clusters, consistent with previous studies.}
    \label{fig:myfigure-UMAP}
\end{figure}

The UMAP projection with \texttt{n\_neighbors} = 6 and the HDBSCAN clustering of repeating FRBs from FRB 20220912 identified three distinct clusters, as shown in Figure \ref{fig:myfigure-UMAP}, each in a different color. 
We used eight parameters in the UMAP machine learning technique to identify the different clusters and investigate the physical properties within each cluster. 
The average value of each parameter in each cluster is shown in Table~\ref{tab:my-table-quantitative}. 
As observed, the three clusters have different average peak frequencies: Cluster 1 at 1082 MHz, Cluster 2 at 1399 MHz (the highest), and Cluster 3 at 1192 MHz. 
This frequency-based clustering indicates that bursts from the same repeating source can vary significantly in their spectral properties.

Additionally, we provide a qualitative description of each parameter in each cluster, as shown in Table~\ref{tab:my-table-qualitative}. 
Notably, Cluster~3 exhibits broadband characteristics and high fluence, features often associated with non-repeating FRBs \citep[e.g.,][]{2021ApJ...923....1P, 2021ApJS..257...59C}. 

\begin{figure*}[htbp]
    \centering
    \includegraphics[width=0.9\textwidth]{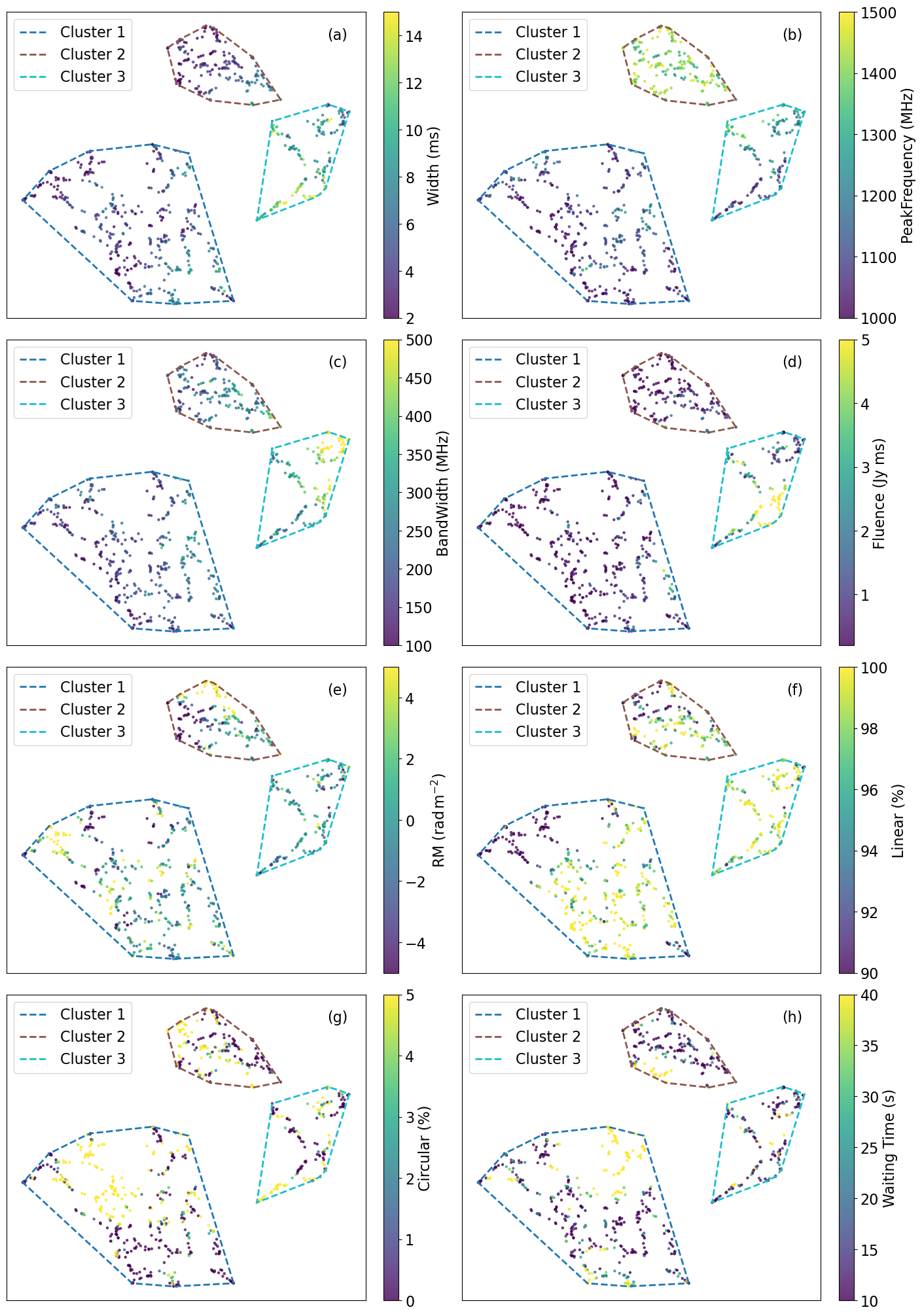}
    \caption{Parameter coloring of the clustering result for \texttt{n\_neighbors} = 6.}
    \label{fig:myfigure-parameter-coloring}
\end{figure*}

\begin{figure*}[htbp]
    \centering
    \includegraphics[width=0.9\textwidth]{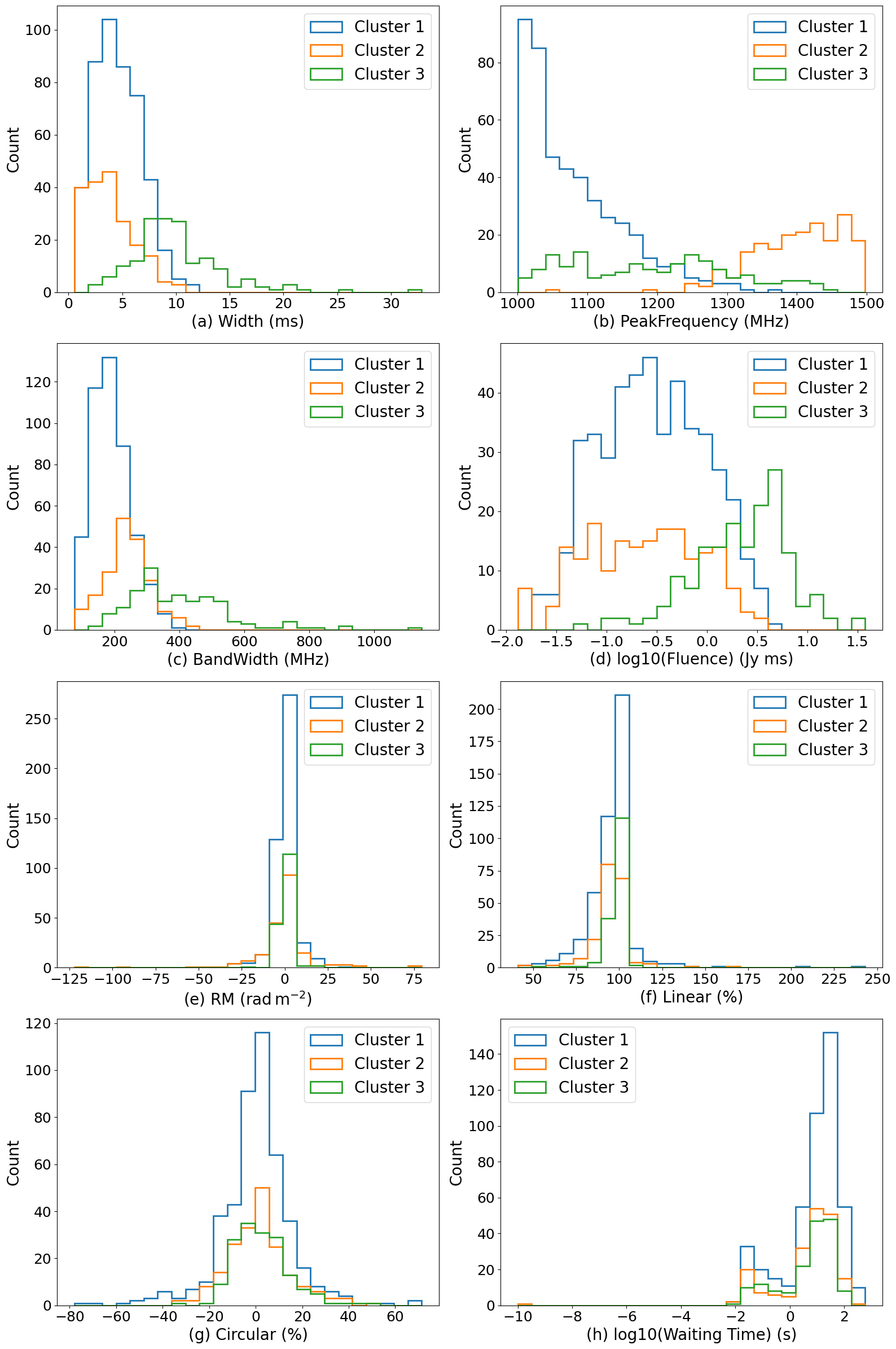}
    \caption{Histogram for each parameter for \texttt{n\_neighbors} = 6.}
    \label{fig:myfigure_histogram}
\end{figure*}

To examine the dominant parameter in the UMAP projection, we also plot the parameter coloring of each cluster, as shown in Figure~\ref{fig:myfigure-parameter-coloring}. 
In Figure~\ref{fig:myfigure-parameter-coloring}(b), we observe that the peak frequency parameter varies significantly across the clusters, which is consistent with the data presented in Table~\ref{tab:my-table-quantitative} and Table~\ref{tab:my-table-qualitative}. 
We also present histograms of each parameter in each cluster (see Fig.~\ref{fig:myfigure_histogram}). 
These findings not only highlight the spectral diversity within a single repeating source but also suggest that machine learning-based clustering can serve as a powerful tool for identifying intrinsic emission modes. 
Further exploration of such methods may lead to deeper insights into the physical origins and classification of FRBs.

\begin{figure}[htbp]
    \centering
    \includegraphics[width=\textwidth]{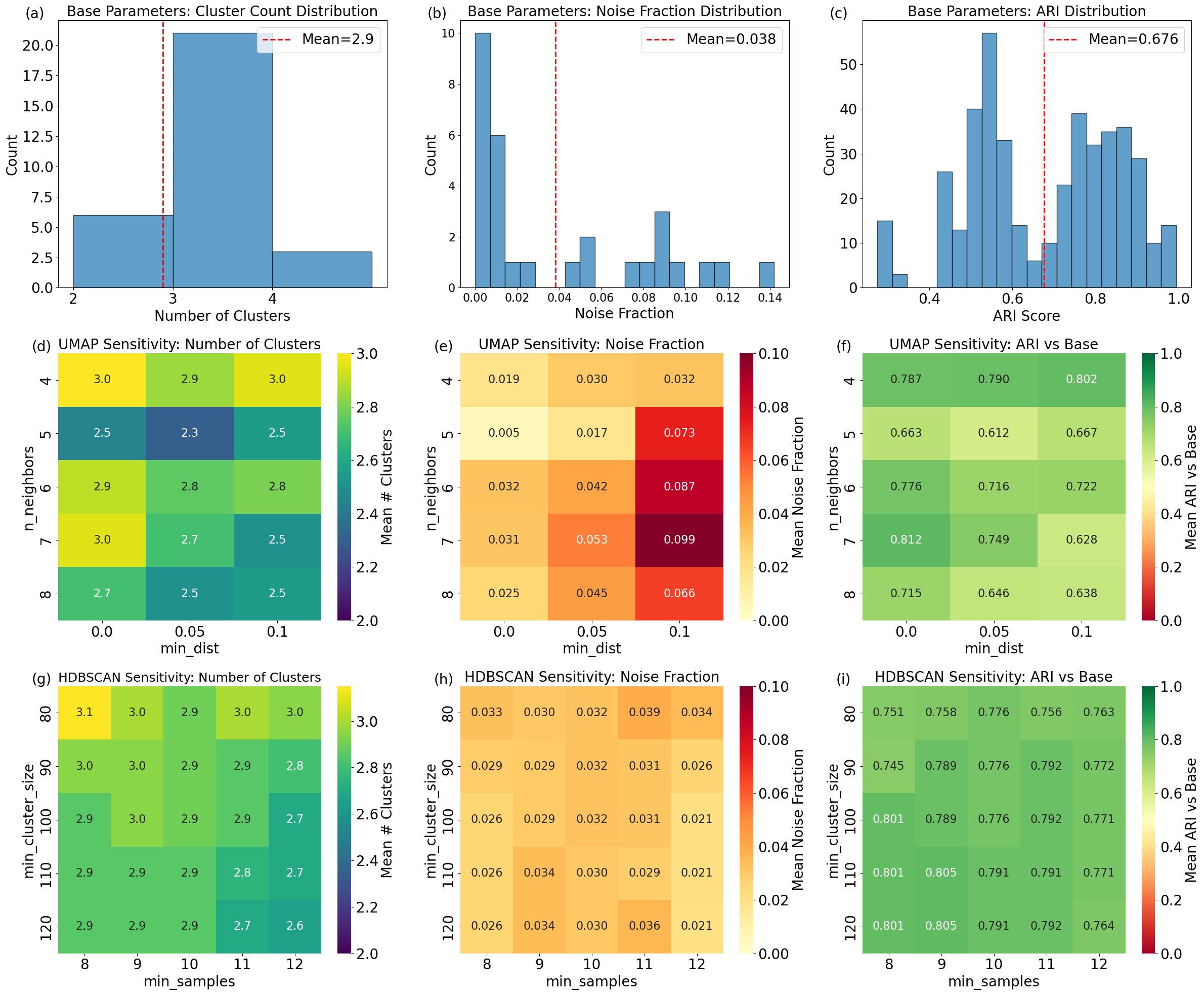}
   \caption{The histograms (a)–(c) show the distributions of cluster count, noise fraction, and ARI, along with their mean values obtained from 30 random seeds under the base hyperparameter settings. The heatmaps (d)–(f) show the UMAP sensitivity check results for \texttt{n\_neighbors} ranging from 4 to 8 and \texttt{min\_dist} ranging from 0 to 0.1, including cluster count, noise fraction, and ARI scores relative to the base results. The heatmaps (g)–(i) show the HDBSCAN sensitivity check results for \texttt{min\_cluster\_size} ranging from 80 to 120 and \texttt{min\_samples} ranging from 8 to 12, including cluster count, noise fraction, and ARI scores relative to the base results.}
    \label{fig:sensitivity analysis}
\end{figure}

\textbf{Hyperparameter Sensitivity Analysis (see Fig.~\ref{fig:sensitivity analysis}):} The baseline hyperparameters (\texttt{n\_neighbors} = 6, \texttt{min\_dist} = 0, \texttt{min\_cluster\_size} = 100, \texttt{min\_samples} = 10) exhibited moderate stability across 30 random seeds, with an average of 2.90 ± 0.54 clusters, a noise fraction of 0.038 ± 0.043, and an ARI of 0.676 ± 0.175. For UMAP, varying \texttt{n\_neighbors} and \texttt{min\_dist} revealed a significant effect on clustering results. Across 30 seeds per parameter combination, ARI values ranged from 0.612 to 0.812, indicating that the results are moderately consistent and stable with the baseline clustering. For HDBSCAN, varying \texttt{min\_cluster\_size} and \texttt{min\_samples} had a smaller effect, with ARI values ranging from 0.745 to 0.805. Most parameter combinations maintained the number of clusters and noise fraction close to the baseline, indicating relatively robust clustering outcomes under small perturbations of HDBSCAN parameters. Overall, the sensitivity analysis demonstrates that the UMAP parameter \texttt{n\_neighbors} = 6, selected based on the \texttt{Silhouette Score}, is locally stable, while HDBSCAN parameters are less sensitive to small changes.

\begin{figure}[htbp]
    \centering    \includegraphics[width=\textwidth]{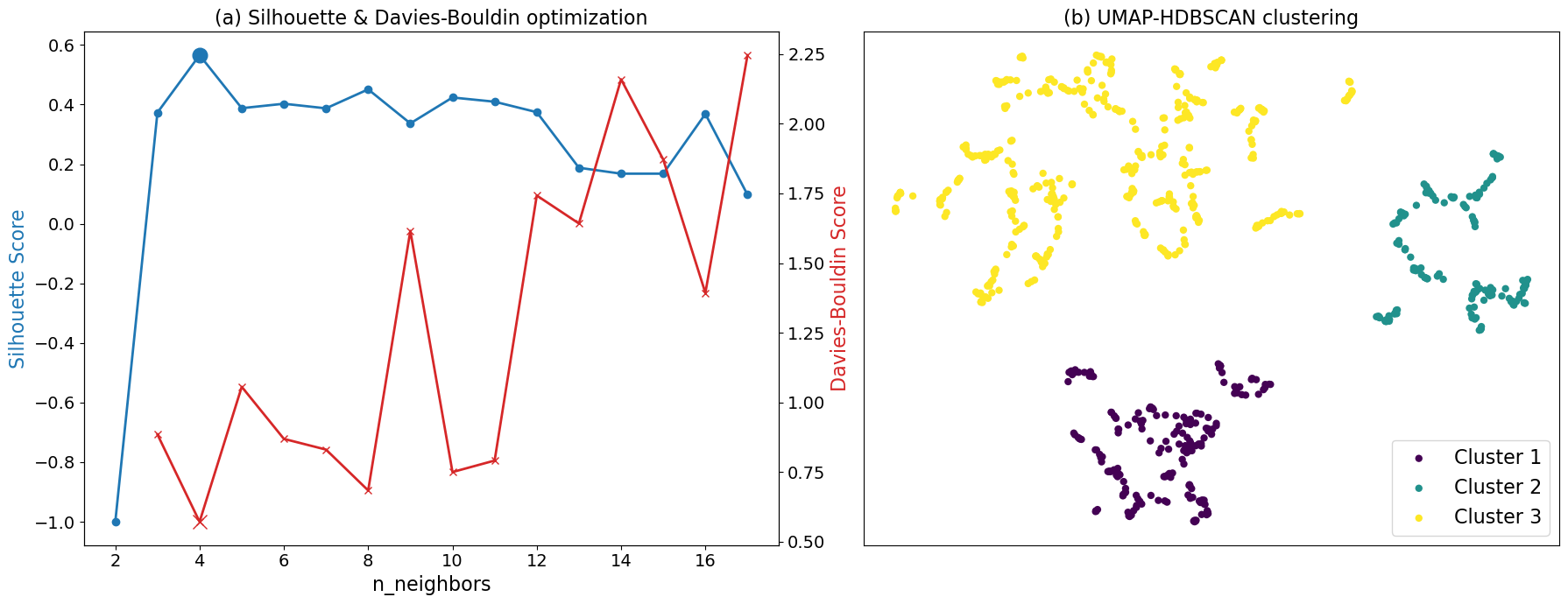}
   \caption{(a) \texttt{Silhouette} and \texttt{Davies-Bouldin Scores} for different \texttt{n\_neighbors} values, calculated without including the Waiting Time feature. The \texttt{Silhouette Score} is maximized and the \texttt{Davies-Bouldin Score} is minimized at \texttt{n\_neighbors = 4}, which was therefore selected for the analysis. (b) UMAP-HDBSCAN clustering of FRB 20220912A, performed without the Waiting Time feature, showing three clusters.}
    \label{fig:no wating time}
\end{figure}

\textbf{Analysis Without Waiting Time (see Fig.~\ref{fig:no wating time}):} Following dimensionality reduction and HDBSCAN clustering without the Waiting Time feature, three clusters were obtained, consistent with the results of the original analysis. This confirms that the identified taxonomy is robust and not driven by observational cadence.

\begin{figure}[htbp]
    \centering
    \includegraphics[width=\textwidth]{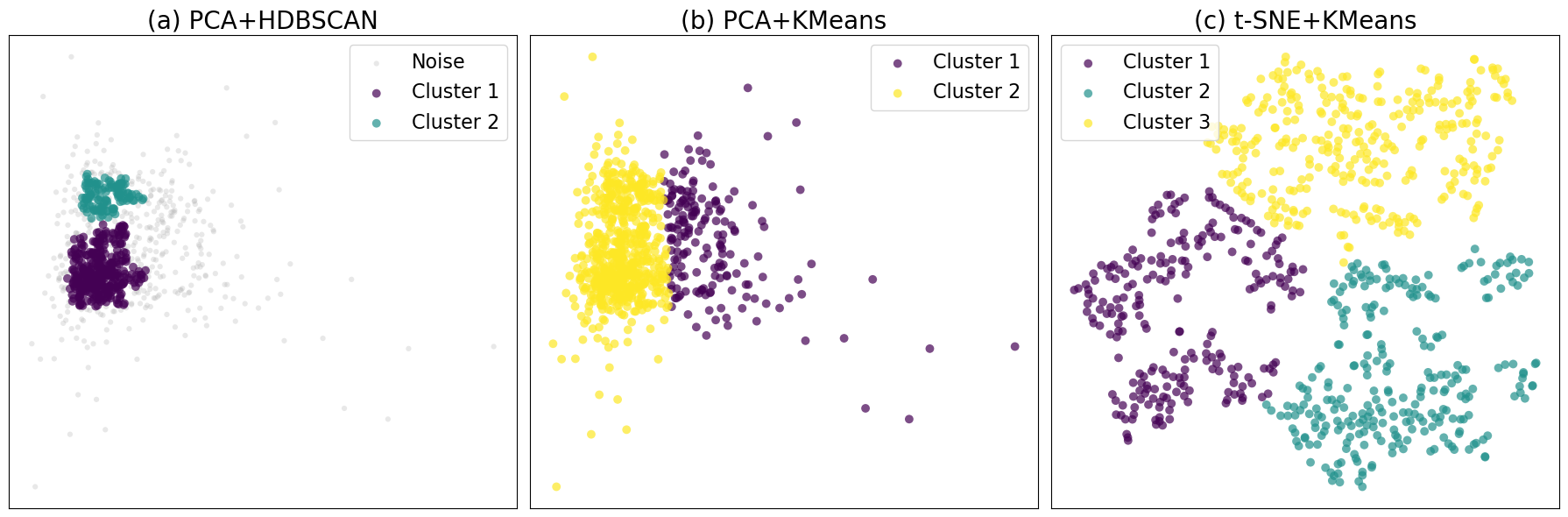}
   \caption{Clustering results using four alternative dimensionality reduction and clustering methods: (a) PCA + KMeans, (b) PCA + HDBSCAN, and (c) t-SNE + KMeans. Among the three cases, there are two clusters in (a) and (b), and three clusters in (c).}
    \label{fig:different method}
\end{figure}

\textbf{Alternative Dimensionality Reduction and Clustering Methods (see Fig.~\ref{fig:different method}):} Across the three alternative methods, the first two methods yielded two clusters each, while the third method produced three clusters. This discrepancy likely arises from the differing characteristics of these dimensionality reduction methods. PCA is a linear technique that projects data onto directions of maximum variance and may fail to separate clusters if the underlying separation is nonlinear. Notably, t-SNE followed by KMeans resulted in three clusters, consistent with the results obtained using our method (UMAP+HDBSCAN), supporting the validity of our approach. Further comparison between the clustering results shows that around 70\% of t-SNE+KMeans Cluster 1 originates from UMAP+HDBSCAN Cluster 3, around 70\% of t-SNE+KMeans Cluster 2 from UMAP+HDBSCAN Cluster 2, and 100\% of t-SNE+KMeans Cluster 3 from UMAP+HDBSCAN Cluster 1, suggesting a substantial overlap between the clustering structures revealed by the two methods. Moreover, from Figs.~\ref{fig:myfigure-Silhoutte-Score} and ~\ref{fig:different method}, the \texttt{Silhouette Score} \citep{ROUSSEEUW198753} suggests that, under the optimal settings, UMAP+HDBSCAN can achieve better cluster separation and compactness than t-SNE+KMeans under the same dataset and clustering evaluation criterion.

\begin{table*}[]
\caption{Average value of each parameter in each cluster with \texttt{n\_neighbors} = 6. The errors include two significant figures.} 
\label{tab:my-table-quantitative}
\begin{tabular}{|c|c|c|c|}
\hline
\shortstack[c]{Average value \\ of each parameter \\ in each cluster} & Cluster 1 & Cluster 2 & Cluster 3 \\
\hline
Peak Frequency (MHz)         & 1082 ± 71     & 1399 ± 67     & 1192 ± 110     \\ \hline
Bandwidth (MHz)              & 190 ± 58      & 240 ± 73      & 397 ± 160         \\ \hline
Fluence (Jy ms)              & 0.56 ± 0.70    & 0.49 ± 0.62    & 3.7 ± 4.6      \\ \hline
Width (ms)                   & 4.6 ± 2.1      & 3.9 ± 2.2      & 10.0 ± 4.3      \\ \hline
RM (rad m$^{-2}$)            & 0.2 ± 6.2      & -1 ± 18       & -0.1 ± 3.7     \\ \hline
Linear (\%)                  & 96 ± 14     & 95 ± 12     & 97.7 ± 5.2     \\ \hline
Circular (\%)                & 0 ± 16      & 2 ± 13      & 2 ± 12      \\ \hline
Waiting Time (s)             & 30 ± 51       & 19 ± 28       & 16 ± 19  
\\ \hline

\end{tabular}
\end{table*}

\begin{table}[]
\caption{The qualitative description with \texttt{n\_neighbors} = 6. The qualitative description of the clusters is based on the range of values for each parameter of a given cluster (except for Linear, Circular, and RM).}
\label{tab:my-table-qualitative}

\begin{tabular}{|c|c|c|c|}
\hline
Parameter     & Cluster 1       & Cluster 2 & Cluster 3       \\ \hline
PeakFrequency & Low             & High      & Moderate        \\ \hline
BandWidth     & Narrowband      & Midband   & Broadband       \\ \hline
Fluence       & Low             & Low       & High            \\ \hline
Width         & Moderate        & Short     & Long            \\ \hline
RM            & Slight Positive & Negative  & Slight Negative \\ \hline
Linear        & High            & High      & Very High       \\ \hline
Circular      & Very Low        & Low       & Low             \\ \hline
Waiting Time   & Long            & Moderate  & Short           \\ \hline
\end{tabular}

\end{table}

\begin{figure}[htbp]
    \centering
    \includegraphics[width=\textwidth]{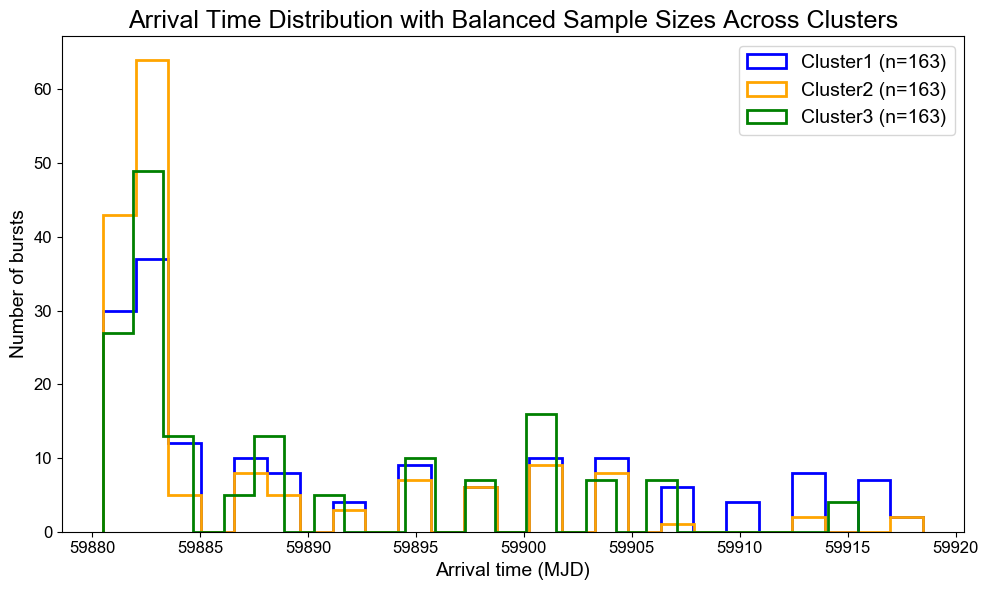}
   \caption{Arrival time distribution for the three clusters. To ensure a fair comparison, each cluster was downsampled to match the sample size of the smallest cluster (Cluster 3 (H) with 163 samples), resulting in balanced sample sizes across clusters.}
    \label{fig:arrival time}
\end{figure}

\section{Discussion} \label{sec:discussion}

\subsection{Similarities with Clusters Identified in \cite{2023MNRAS.521.5738C} and \cite{2023MNRAS.524.1668R}}

In recent years, unsupervised machine learning techniques have been increasingly used to classify repeating FRBs and uncover subtle structures in their observed properties.
\cite{2023MNRAS.521.5738C} applied UMAP (Uniform Manifold Approximation and Projection) followed by K-means clustering to analyze bursts from FRB 20201124A, using data from the Five-hundred-meter Aperture Spherical Radio Telescope (FAST). They identified three distinct clusters of bursts: Cluster 1 (C) characterized by high peak frequency, broad bandwidth, and low fluence; Cluster 2 (C) with low peak frequency, narrow bandwidth, and low fluence; and Cluster 3 (C) with high peak frequency, broad bandwidth, and high fluence. This approach expands upon previous classification methods based on Waiting Time, and their results suggest a new categorization scheme that may reflect intrinsic differences in burst generation mechanisms. Notably, they observed similar clustering behavior in FRB 121102, implying a possible shared physical origin for these frequently repeating sources.

\cite{2023MNRAS.524.1668R} also utilized UMAP in combination with HDBSCAN (Hierarchical Density-Based Spatial Clustering of Applications with Noise) to classify 1652 bursts from FRB 121102, also detected by FAST. Their analysis revealed three physically distinct clusters within 1613 valid bursts: Cluster 1 (R), characterized by narrow bandwidth and low fluence; Cluster 2 (R), broadband with diverse fluence; and Cluster 3 (R), narrowband with high fluence. These clusters exhibited diverse properties and suggested different origins, emphasizing the heterogeneity within the FRB population. Their approach also employed a broader set of parameters than previous studies, offering a more refined and comprehensive classification framework. To qualitatively compare our clustering results with those of these two studies, we present Tables~\ref{tab:my-table-qualitative-Bo} and~\ref{tab:my-table-qualitative-BJ}, which summarize the key morphological properties of each cluster across different sources and studies.

\begin{table*}[]
\caption{This qualitative table compares our results with those from \cite{2023MNRAS.521.5738C}. The red words highlight the inconsistencies between our results and those from \cite{2023MNRAS.521.5738C}. (H): This work; (C): \cite{2023MNRAS.521.5738C}.}
\label{tab:my-table-qualitative-Bo}
\begin{flushleft}
\begin{tabular}{|c|
c 
c |
c 
c |
c 
c |}
\hline
 & FRB 20220912A & FRB 20201124A & FRB 20220912A & FRB 20201124A & FRB 20220912A & FRB 20201124A \\ \hline
Cluster ID          & Cluster 1 (H) & Cluster 2 (C) & Cluster 2 (H)                  & Cluster 1 (C)                    & Cluster 3 (H)                   & Cluster 3 (C)               \\ \hline
PeakFrequency & Low           & Low           & High                           & High                             & {Moderate} & {High} \\ \hline
BandWidth     & Narrowband    & Narrowband    & {Midband} & {Broadband} & Broadband                       & Broadband                   \\ \hline
Fluence     & Low           & Low           & Low                            & Low                              & High                            & High                        \\ \hline
\end{tabular}
\end{flushleft}
\end{table*}

\begin{table*}[]
\caption{This qualitative table compares our results with those from \cite{2023MNRAS.524.1668R}. The red words highlight the inconsistencies between our results and those from \cite{2023MNRAS.524.1668R}. (H): This work; (R): \cite{2023MNRAS.524.1668R}.}
\label{tab:my-table-qualitative-BJ}
\begin{flushleft}
\begin{tabular}{|c|
>{}c 
>{}c |
>{}c 
>{}c |
>{}c 
>{}c |}
\hline
 & FRB 20220912A & FRB121102 & FRB 20220912A & FRB121102 & FRB 20220912A & FRB121102 \\ \hline
Cluster ID      & Cluster 1 (H) & Cluster 1 (R) & Cluster 2 (H)                  & Cluster 3 (R)                     & Cluster 3 (H)               & Cluster 2 (R)                  \\ \hline
BandWidth      & Narrowband    & Narrowband    & {Midband} & {Narrowband} & Broadband                   & Broadband                      \\ \hline
Fluence/Energy & Low           & Low           & {Low}     & {High}       & {High} & {Diverse} \\ \hline
\end{tabular}
\end{flushleft}
\end{table*}

\begin{table}[]
\caption{This quantitative table compares our results with those from \cite{2023MNRAS.521.5738C}. (H): This work; (C): \cite{2023MNRAS.521.5738C}. The errors include two significant figures.}
\label{tab:my-table-quantitative-Bo}
\setlength{\tabcolsep}{4pt}
\begin{flushleft}
\begin{tabular}{|c|
>{}c 
>{}c |
>{}c 
>{}c |
>{}c 
>{}c |}
\hline
 & FRB 20220912A & FRB 20201124A & FRB 20220912A & FRB 20201124A & FRB 20220912A & FRB 20201124A \\ \hline
\shortstack[c]{Average value \\ of each parameter \\ in each cluster} & Cluster 1 (H) & Cluster 2 (C) & Cluster 2 (H)                  & Cluster 1 (C)                    & Cluster 3 (H)                   & Cluster 3 (C)               \\ \hline
Peak Frequency (MHz) & 1082 ± 71            & 1038 ± 34           & 1399 ± 67                          & 1259 ± 140                             & 1192 ± 110 & 1157 ± 120 \\ \hline
Bandwidth (MHz)     & 190 ± 58    & 137 ± 54    & 240 ± 73 & 311 ± 93 & 397 ± 160                       & 273 ± 100                   \\ \hline
Fluence (Jy ms)      & 0.56 ± 0.70           & 0.947 ± 0.034          & 0.49 ± 0.62                           & 0.9 ± 1.0                              & 3.7 ± 4.4                            & 6.0 ± 6.0                        \\ \hline
\end{tabular}
\end{flushleft}
\end{table}

For clarity, we use the same notation as in Table~\ref{tab:my-table-qualitative-Bo}: (H) denotes the results of this work, while (C) represents the values from \cite{2023MNRAS.521.5738C}. In this work, we identify three distinct clusters: Cluster 1 (H), with low peak frequency, narrow bandwidth, and low fluence; Cluster 2 (H), with high peak frequency, mid-range bandwidth, and low fluence; and Cluster 3 (H), with moderate peak frequency, broad bandwidth, and high fluence. We compare our results with previous studies by \citet{2023MNRAS.521.5738C} and \citet{2023MNRAS.524.1668R}, who also employed unsupervised machine learning techniques to classify repeating FRBs.

As shown in Table~\ref{tab:my-table-qualitative-Bo}, one of the clusters we identified—Cluster 1 (H), characterized by low peak frequency, narrow bandwidth, and low fluence—closely resembles the "Cluster 2 (C)" bursts reported by \citet{2023MNRAS.521.5738C} for FRB 20201124A. Similarly, Table~\ref{tab:my-table-qualitative-BJ} shows that our Cluster 1 (H) is comparable to Cluster 1 (R), as both exhibit narrow bandwidth; in our case, it is also associated with low fluence, while theirs is associated with low energy. These similarities suggest a potential correspondence between our clustering results and those derived from other repeating sources.
 
These narrow bandwidth and low fluence properties of the burst correspond to the typical characteristics of repeating FRBs \citep{2021ApJ...923....1P}.
Notably, Cluster 3 (H) and Cluster 3 (C) both exhibit broadband characteristics and high fluence, which implies higher burst energy.

Apart from Table~\ref{tab:my-table-qualitative-Bo}, we also provide Table~\ref{tab:my-table-quantitative-Bo}, which summarizes a quantitative comparison between our clustering results and those reported by \cite{2023MNRAS.521.5738C}. 
The redshifts of FRB 20220912A and FRB 20201124A are 0.0771 \citep{2023ApJ...949L...3R} and 0.098±0.002 \citep{2021ATel14516....1K}, respectively. 
Therefore, the redshift effects, including the frequency shift, are negligibly small compared with the error values in Table \ref{tab:my-table-quantitative-Bo}.
The listed uncertainties are given to two significant figures.

Overall, the clustering analysis reveals consistent trends between the two studies. In particular, our Cluster 1 (H) and Chen’s Cluster 2 (C) both exhibit similarly low average peak frequencies. Likewise, the average peak frequencies of our Cluster 2 (H) and Chen’s Cluster 1 (C) are consistent within the measurement errors. 
Cluster 3 in both works lies in between, with average peak frequencies around 1190–1160 MHz.

For bandwidth, \cite{2023MNRAS.521.5738C} reports a broader spread in Cluster 1 (C) and Cluster 3 (C), while our results for Cluster 3 (H) also indicate a large average bandwidth (around 400 MHz). Regarding fluence, both works display large dispersions, and the average fluence of our Cluster 3 (H) is consistent with Chen’s Cluster 3 (C) within the measurement errors.

These comparisons suggest that the overall clustering structure is similar between FRB 20220912A and FRB 20201124A. The average values of the peak frequency, bandwidth, and fluence in our clusters are consistent with those reported by \cite{2023MNRAS.521.5738C} within the measurement errors, indicating overall agreement between the clusters in the two FRB sources.


\subsection{Physical Interpretation of the Observed Clusters}

We identify three distinct clusters, consistent with the clustering reported by \cite{2023MNRAS.521.5738C} for FRB 20201124A (see Table~\ref{tab:my-table-quantitative-Bo} for comparison). As shown in Figure~\ref{fig:arrival time}, the histogram reveals no distinct differences in arrival times among the clusters. The bursts span a period of approximately 30 days, and for any given day, the arrival time distributions of the different clusters appear nearly identical. This lack of temporal segregation makes the interpretation based on time evolution less compelling in our case.

The three distinct clusters with distinct physical properties 
can be difficult to reconcile with synchrotron maser models, which predict that the emission frequency depends on the upstream plasma density and shock Lorentz factor \citep{2014MNRAS.442L...9L, 2017ApJ...843L..26B, 2019MNRAS.485.4091M}. To explain the three distinct clusters within this framework, the maser scenario would require 
fine-tuned or clustered plasma densities or Lorentz factors to reproduce the distinct peak frequency clusters, all while operating under similar temporal conditions, given that the arrival times do not differ significantly across clusters. For instance, while frequency clustering on day scales has been observed in the periodically modulated FRB 20180916B \citep{2021ApJ...911L...3P,2021Natur.596..505P}, Fig.~\ref{fig:arrival time} shows no such indication. Thus, while not ruled out, such maser scenarios appear difficult to reconcile without invoking additional relatively small-scale structure or fine-tuning.


Alternatively, in pulsar-like models, curvature radiation from charged particle bunches traveling along curved magnetic field lines can produce coherent radio emission \citep[e.g.,][]{2012MNRAS.423.2464W}. While such radiation typically yields a broad spectrum \citep{2018ApJ...868...31Y}, a more plausible explanation for the observed clustering is that different frequencies originate at different heights along the magnetic field lines. Subtle changes in the observer’s line of sight relative to the magnetar’s magnetic field lines could then naturally produce distinct frequency clusters on short timescales, similar to the precession model for the chromatic active window of periodically modulated FRBs \citep{2021ApJ...909L..25L}. In this context, Cluster 2 exhibits the highest average peak frequency, followed by Cluster 3, and then Cluster 1 (see Tables~\ref{tab:my-table-quantitative} and~\ref{tab:my-table-qualitative}).
\cite{2021ApJ...909L..25L} presented the curvature radiation models to showcase three different frequencies of FRBs, which correspond to three different altitudes.
Therefore, Cluster 2 would have the lowest altitude of burst emission, whereas Cluster 3 and 1 would be at higher altitudes. This scenario is consistent with the absence of arrival-time clustering on day scales (Fig.~\ref{fig:arrival time}).
A more detailed temporal analysis of these clusters will be feasible in the future once larger burst samples are available.

\subsection{Broadband Characteristics and Similarities with Non-Repeating FRBs}

One major challenge in FRB classification is the potential misclassification between repeaters and non-repeaters. Due to observational limitations such as limited sensitivity, sparse monitoring cadence, and large distances to host galaxies, some repeating FRBs may be detected only once and thus misidentified as non-repeaters \citep{2024MNRAS.52711158Y,2025arXiv250609138B}. This observation suggests that a subset of so-called non-repeating FRBs might actually represent the brighter portion of repeating sources.

Therefore, it is valuable to compare the physical properties of bright repeating bursts with those of non-repeating FRBs. The broadband and high-fluence nature of bursts in Cluster 3 (H) and Cluster 3 (C) is particularly noteworthy, as these characteristics closely resemble those typically observed in non-repeating FRBs. Furthermore, as shown in Table~\ref{tab:my-table-qualitative-BJ}, Cluster 1 (H) is consistent with Cluster 1 (R), which is characterized by narrow bandwidth and low fluence/energy—features that are commonly associated with repeating FRBs \citep{2021ApJ...923....1P}. This comparison highlights the spectrum of burst morphologies within repeating sources and raises the possibility that the bright broadband bursts from repeaters may mimic the appearance of non-repeating FRBs.

To further understand the nature of these clusters, future work could incorporate additional parameters, such as dispersion measure, polarization properties, and temporal evolution. These features may help distinguish whether the observed frequency diversity arises from changes in the physical state of the source or from external propagation effects. Moreover, comparing our findings with other well-characterized repeaters, such as FRB121102 and FRB 20190520B, could help determine whether frequency-based clustering is a general feature among repeating FRBs or a unique signature of FRB 20220912A.

\section{Conclusion} \label{sec:conclusion}

The classification and interpretation of fast radio bursts (FRBs) present a significant challenge due to their short durations, high variability, and the growing volume of observed events. Machine learning techniques offer a powerful, unbiased framework to simultaneously analyze multiple physical parameters and uncover hidden structures within FRB populations. In particular, unsupervised learning methods have shown great potential in identifying subtypes of FRBs that may reflect different physical origins.

In this work, we applied unsupervised machine learning techniques—UMAP and HDBSCAN—to analyze eight observational parameters of bursts from FRB 20220912A. The analysis identified three distinct clusters with varying spectral and fluence properties.

By comparing our results with previous clustering studies on other repeaters such as FRB 20201124A and FRB121102, we find that some of our clusters exhibit similar characteristics, suggesting potentially shared emission mechanisms across different sources. We also provide qualitative physical interpretations for each cluster, offering insights into the spectral diversity within a single repeating source.

In addition, we performed a series of robustness tests, including hyperparameter sensitivity checks and clustering without the Waiting Time feature, which consistently support the three-cluster structure obtained from UMAP+HDBSCAN. When using alternative combinations of dimensionality-reduction and clustering methods, we found that PCA generally did not capture three clusters, whereas t-SNE+KMeans analysis successfully yielded three clusters, consistent with the UMAP+HDBSCAN result.

The lack of significant temporal differences among clusters suggests that time evolution alone cannot explain the observed diversity. Notably, the distinct clusters, each exhibiting different physical properties such as peak frequency and burst width, may not be straightforward to reconcile with synchrotron maser. 
A more plausible explanation involves pulsar-like curvature radiation, where different frequencies originate at different heights along magnetic field lines, and slight changes in the observer’s line of sight could produce distinct frequency clusters.

Moreover, one of the clusters shows broadband emission and high fluence, which are typically associated with non-repeating FRBs. This finding raises the possibility that some non-repeating FRBs may represent the bright end of a repeating source population, possibly misclassified due to observational limitations.

Overall, our results highlight the utility of machine learning techniques in revealing the intrinsic diversity of FRB emission properties, and pave the way for future studies aimed at understanding the physical origins and classification of FRBs.

\section{Acknowledgments}
The authors thank the referee for their insightful comments, which significantly improved the quality of this manuscript. 
A.-C.H acknowledges support from the National Chung Hsing University through the Undergraduate Research Project Scholarship (grant no. 11428535H). 
TH acknowledges the support from the NSTC through grants 113-2112-M-005-009-MY3, 113-2123-M-001-008-, 111-2112-M-005-018-MY3, and the Ministry of Education of Taiwan through a grant 113RD109.
T.W. is supported by Grants-in-Aid for Scientific Research, Nos. JP25KJ0024, JP25K17378 from the Ministry of Education, Culture, Sports, Science and Technology (MEXT) of Japan.
We are also grateful to our collaborators in the NTHU \& NCHU Cosmology Group, especially Dr. Shotaro Yamasaki, a research scholar at NCHU, and Simon C.-C. Ho, a PhD student at the Australian National University, for their many insightful comments and for their continuous support and invaluable contributions throughout this project.

\section{Data availability}
The data underlying this article is available in the work of \cite{2023ApJ...955..142Z} . Other data described in this article will be shared upon reasonable request to the corresponding author.

\bibliography{PASPsample631}{}
\bibliographystyle{aasjournal}



\end{document}